\documentstyle[aps,prl,preprint]{revtex}

\begin{document}

\draft
\title{Universality of Random-Matrix Results for non-Gaussian Ensembles}
\author{G. Hackenbroich and H.A. Weidenm\"uller}
\address{Max-Planck-Institut f\"ur Kernphysik, 69117 Heidelberg, Germany}
\date{\today}
\maketitle

\begin{abstract}
We study random-matrix ensembles with a non-Gaussian probability distribution
$P(H) \sim \exp (-N {\rm tr }\, V(H))$ where $N$ is the dimension of the
matrix $H$ and $V(H)$ is independent of $N$. Using Efetov's supersymmetry
formalism, we show that in the limit
$N \rightarrow \infty$ both energy level correlation functions and
correlation  functions of $S$-matrix elements are independent of $P(H)$
and hence universal on the scale of the local mean level spacing.
This statement applies to each of the three generic ensembles (unitary,
orthogonal, and  symplectic). Universality is also found for
correlation functions depending on some external parameter. Our results
generalize previous work by Brezin and Zee [Nucl.\ Phys.\ B {\bf 402},
613 (1993)].
\end{abstract}


The energy levels of a variety of physical systems including complex
nuclei, disordered conductors and classically chaotic systems exhibit
universal behaviour: The
statistical properties of the spectrum depend
only on the fundamental symmetries of the underlying Hamiltonian and
can be
described by random-matrix theory \cite{BM}. Three symmetry classes exist:
Systems with broken time-reversal symmetry are described by the
unitary ensemble, time-reversal invariant systems by the symplectic
or the orthogonal ensemble depending on whether spin-orbit coupling is
present or not. The symmetry of the Hamiltonian does
not specify the random-matrix ensemble completely. In general, one
requires in addition that the matrix elements be
statistically independent from each other. This additional condition
restricts the probability density for all three ensembles to a Gaussian
form and defines the Gaussian matrix ensembles. However, the assumption of
statistical independence is used mainly for the sake of mathematical
simplicity and is not motivated by first principles. Moreover, while
accounting for the local energy level correlations on the scale of
the mean level spacing $d$, the Gaussian ensembles fail to describe the
global properties of the experimentally observed spectra. In particular,
they predict a universal form for the mean density $\rho(E)$
of eigenvalues (the semi-circle law for systems
with a large number of levels), whereas this quantity is known to be system
specific and non-universal.

Within random-matrix theory, more realistic forms of $\rho(E)$ are obtained
when one considers non-Gaussian ensembles. Based on numerical studies of such
ensembles \cite{numerics} it has been conjectured \cite{BM} that their
{\em local}
spectral properties are independent of the measure, identical to those of the
Gaussian
ensembles and hence universal. The effect of a non-Gaussian measure on the
spectral statistics was first analytically investigated
by Brezin and Zee \cite{Brezin2}. Within the unitary ensemble
these authors proved universality for the local two-point level correlation
function. (Properly smoothed level correlations on scales large compared to $d$
were found to depend on the
measure only through the endpoints of the spectrum. This latter
result was later generalized
to all three ensembles by Beenakker \cite{Beenakker}.)

In the present paper, we prove the universality of arbitrary
local correlation functions for any of the three generic matrix ensembles in
the limit $N \rightarrow \infty$, where $N$ denotes the number of levels.
More specifically, we consider a correlation function
$C$ involving an arbitrary number of level-density factors and/or $S$-matrix
elements. This function may depend parametrically on energy arguments and/or
on additional parameters like the strength of an external magnetic field. We
do allow for symmetry breaking. (An external magnetic field, for instance,
breaks orthogonal symmetry). We compare the correlation function $C_G$
evaluated for the Gaussian ensemble, and its analogue $C_P$ evaluated for a
non-Gaussian ensemble having the same symmetry and a distribution $P(H)$
defined as in Eq.\ (\ref{pdensity}) below.
Under the single assumption that all parameters in $C$ range over an interval
$\Delta$ containing on average a finite number of levels $\Delta
\sim {\cal O}(N^{-1})$,
we show that $C_G(d_G(E))=C_P(d_P(E))$. Here, $d_G(E)$ and $d_P(E)$ are the
average local mean level spacings for the Gaussian and the non-Gaussian
ensemble, respectively, both evaluated at energy $E$, which is located at the
center of the interval $\Delta$. These very general results are obtained by
using the supersymmetry method. To allow for non-Gaussian
probability measures, our derivation differs from the usual
formulation of this method \cite{Efetov,blue},

For definiteness we consider the unitary ensemble in the following.
We emphasize that the orthogonal and the symplectic cases can be
treated along exactly parallel lines. Accordingly, we study an
ensemble of $N \times N$ Hermitian random matrices $H$
with volume element  $d [ H ] =
\prod_{i \geq j}^{N} d {\rm Re} H_{ij} \, \prod_{i>j}^N d {\rm Im}
H_{ij}$. The probability density $P(H)$ is defined by
\begin{equation}
  P(H)=Z^{-1} \exp \{ -N \, {\rm tr}\, V(H) \},
\label{pdensity}
\end{equation}
where $Z$ is a normalization constant. This is the most general density
compatible with the basic assumption of random-matrix theory, namely
that $P(H)dH$ is invariant under unitary transformations $H \rightarrow
U H U^{-1}$. The function $V$ is assumed both to confine the spectrum to some
finite interval and to generate a smooth mean level
density, in the limit
$N \rightarrow \infty$ \cite{Kravtsov}. Then, for the ensemble defined by Eq.\
(\ref{pdensity}), the mean level spacing $d$ is of order $N^{-1}$. Note,
that $V(H)=g H^2$, $g>0$ defines the Gausssian unitary ensemble.

In the supersymmetry method, we generically express correlation
functions as derivatives of a generating functional \cite{blue}. The latter is
written in terms of an integral over a supervector $\Psi$ with bosonic
(commuting) and fermionic (anticommuting) components
\begin{equation}
I=\int d [ \Psi ] \langle \exp \{ {i\over 2} \Psi^{\dagger} {\bf  L}^{1/ 2}
({\bf H}-{\bf E}+{\bf M}){\bf L}^{1/2} \Psi\} \rangle ,
\label{genfunc}
\end{equation}
where the brackets denote the ensemble average $\langle \ldots \rangle
\equiv \int d [ H ] P(H) (\ldots)$. For example, in the case of the 2-point
function at energies $E$ and $E'$ one defines the supervector by
$\Psi^T = (S_1^T,\chi_1^T,S_2^T,\chi_2^T)$ with complex bosonic entries
$S_1$, $S_2$ and complex fermionic entries $\chi_1$, $\chi_2$, each
entry being itself an $N$-dimensional vector. The measure has the form
$d[\Psi]=\prod_{\mu=1}^N \prod_{j=1}^2 idS_{\mu j}^* d
S_{\mu j} d \chi_{\mu j}^* d \chi_{\mu j}$.
The Hamiltonian ${\bf H}$ is the direct product of the $N \times
N$ Hamiltonian $H$ and the unit matrix in the superspace. The energy
${\bf E}$ stands for the product of the mean energy $E=(E_1+E_2)/2$ and the
unit matrix in both the level space and the superspace, ${\bf L}$ is the
direct product of the unit matrix in level space with
$L={\rm diag}(1,1,-1,-1)$.
The matrix ${\bf M}$ contains energy differences and the source terms.
In the case of scattering problems, ${\bf M}$ also
contains couplings to external channels. To account for dependences on
external parameters, ${\bf M}$ may contain additional random matrices besides
$H$ over which additional ensemble averages must be performed. We postpone
this calculation and confine attention to the ensemble average over $H$.
In any case, we have ${\bf M} = O(N^{-1})$
because we are interested in correlations involving energies of the
order of the mean level spacing $d \sim N^{-1}$.
For $n$-point functions with $n > 2$, the {\em form} of Eq.\ (\ref{genfunc})
remains unchanged, although the dimensions of the vectors $\Psi$,
$\Psi^\dagger$ and of the matrices ${\bf H}$, ${\bf E}$, ${\bf L}$, ${\bf M}$
in superspace will increase. Our proof applies to all these cases because
it is independent of these dimensions.

In the
Gaussian case, one usually decouples the interaction generated by the ensemble
average by means of a Hubbard-Stratonovich transformation \cite{blue}. This
procedure introduces a supermatrix $\sigma$ and maps the generating functional
onto
a non-linear $\sigma$-model. The procedure
relies on the Gaussian form of the probability density and does not
apply to general $P(H)$. The central point of our argument is based on the
observation that it is nevertheless possible to introduce the "composite
variables" $\sigma$ for any $P(H)$. Indeed, for any $P(H)$ the unitary
invariance of the ensemble implies that for $N \rightarrow \infty$,
the integrand in Eq.\ (\ref{genfunc}) depends on
$\Psi$ and  $\Psi^\dagger$ only via the invariant form
$A_{\alpha \beta}={1\over N}L_{\alpha \gamma}^{1/2}\sum_{\mu=1}^N
\Psi_{\mu \gamma} \Psi_{\mu \delta}^{\dagger} L_{\gamma \beta}^{1/2}$.
Here $\mu$, $\nu$ are level indices and $\alpha$, $\beta$, $\gamma$, $\delta$
superindices \cite{termm}. We explicitly introduce a
supermatrix $\sigma$ with the same dimension and symmetry properties as $A$
by writing $I$ as an integral over
a $\delta$-function,
\begin{equation}
  I=\int d [ \Psi ] \int d \sigma \, \delta(\sigma -A)
  \langle \exp \{ {i\over 2} \Psi^{\dagger} {\bf L}^{1/2}
  {\bf G} {\bf L}^{1/2} \Psi\} \rangle ,
\label{gensigma}
\end{equation}
with the abbreviation ${\bf G} \equiv {\bf H}-{\bf E}+{\bf M}$. The
$\delta$-function is replaced by its Fourier representation
\begin{eqnarray}
  I= \int d [ \Psi ] \int & & d \sigma
  \, \int d \tau \, \exp \{ {i\over2}N\, {\rm trg} (\tau \sigma) \}
   \nonumber \\
  & & \times \langle \exp \{ {i\over 2} \Psi^{\dagger} {\bf L}^{1/2}
  ({\bf G}-\tau) {\bf L}^{1/2} \Psi\} \rangle ,
\label{genmu}
\end{eqnarray}
and the multiple Gaussian integral over the $\Psi$-supervector is performed
\begin{equation}
  I \! = \! \int\!\!\! d \sigma \!\! \int\!\!\! d \tau \!
  \exp \{ {i\over2}N {\rm trg} (\tau \sigma) \}
  \langle \exp \{ -{1\over 2}\, {\rm tr}\, {\rm trg}\, \ln [
  {\bf G}-\tau] \} \rangle .
\label{gensigmu}
\end{equation}
We have now expressed the functional $I$ as a
integral over two coupled supermatrices $\sigma$ and $\tau$ which
contain all relevant degrees of freedom. In the limit $N \rightarrow \infty$,
the remaining integrals can be done explicitely using the saddle-point
approximation. In particular, it will turn
out that the (diagonal) saddle-point of the $\sigma$-integral determines
the mean level density. These steps will prove the claimed
universality by comparison with the well-known Gaussian case.

To perform the ensemble average we transform $H$ to diagonal form,
$(U H U^{\dagger}) = \Lambda$, and integrate separately over eigenvalues
$\Lambda$ and eigenvectors $U$. Expanding in powers of ${\bf M}$, we have
\begin{eqnarray}
  \lefteqn{ \langle \exp \{ -{1\over 2} {\rm tr} \,
  {\rm trg}\, \ln [ {\bf G}-\tau] \} \rangle
  = \langle \exp \{ -{1\over 2} {\rm tr}\, {\rm trg}\, \ln  D \}  }
  \nonumber \\
  & & \times \exp \{ -{1\over2}\, {\rm tr}\, {\rm trg}\, [
1+\sum_{n=1}^{\infty}
  {(-1)^{n+1}\over n}(D^{-1} U^{\dagger} {\bf M} U))^n ] \}
  \rangle ,
\label{average}
\end{eqnarray}
where $D \equiv (\Lambda -E-\tau)$ is
diagonal in the level space. The
expansion in powers of ${\bf M}$ in Eq.\ (\ref{average}) cannot be
terminated with the first-order term because any power of ${\bf M}$ may
be of the same order $N^{-1}$ as ${\bf M}$. This is the case for $S$-matrix
correlation functions \cite{blue}. The distribution of
eigenvectors does not depend on the form of the probability density
in Eq.\ (\ref{pdensity}). In the large-$N$ limit the eigenvectors
are Gaussian distributed \cite{numerics} and the average over eigenvectors is
evaluated using
Wick contractions. To leading order in powers of $N^{-1}$ we find
that the last exponential in Eq.\ (\ref{average}) takes the simple form
$-{1\over2}\, {\rm tr}\, {\rm trg}\, [ 1+({1\over N} {\rm tr} D^{-1}) {\bf
  M} ]$.
The remaining eigenvalue-integrations are done using the saddle-point
approximation \cite{Brezin1}.
Explicitly, the average over eigenvalues appearing in Eq.\ (\ref{average})
involves the exponential
\begin{eqnarray}
   -{1\over 2} \sum_{\mu} {\rm trg} \ln & &( D)_{\mu}
   -{1\over 2} {\rm tr}\, {\rm trg}\, \ln
   \left[ 1 +\left( {1\over N} \sum_{\mu} (D^{-1})_{\mu}
   \right) {\bf M} \right]
   \nonumber  \\
   & & - N \sum_{\mu}V(\lambda_{\mu}) +2\sum_{\mu <
   \nu} \ln |\lambda_\mu - \lambda_\nu | \, ,
\label{intev}
\end{eqnarray}
where the last term results from the Jacobian associated
with the transformation from the matrix elements of $H$ to
its eigenvalues. Of the four terms in expression (\ref{intev}) the first
is ${\cal O} (N)$ and the second ${\cal O} (1)$. The last two terms are
${\cal O}
(N^2)$ and determine the saddle point values $\lambda_{\mu}^
{\rm sp}$. The calculation is explicitly carried out in ref.\ \cite{Brezin1}
and introduces the average local
level
density $\rho(E)$, and the resolvent $F$, defined by
\begin{equation}
  {1\over N} \sum_{\mu} {1\over E+\tau -\lambda_\mu^{\rm sp}}
  \stackrel{N \rightarrow \infty}{\longrightarrow} \int d E' {\rho (E')\over
  E+\tau -E'} \equiv F(E+\tau).
\end{equation}
Substituting $\lambda_{\mu}^{\rm sp}$ for $\lambda_{\mu}$ in the first two
terms of Eq.\ (\ref{intev}), expanding the terms $\sim {\cal O}(N^2)$
around the saddle-point and performing the Gaussian integrals, we find
\begin{eqnarray}
  I \! = \!&  & \!\int \!\! d \sigma \!\! \int \!\!d \tau \exp \left\{
  {i\over2}N {\rm trg} (\tau
  \sigma) -{1\over 2} \sum_{\mu} {\rm trg} \ln [
  \lambda_{\mu}^{\rm sp}\!\! -E\!\!-\tau ] \right\}\nonumber  \\
  & & \times \exp \left\{ -{1\over 2}\, {\rm tr}\, {\rm trg}\, \ln [
  1+F(E+\tau) {\bf M} ] \right\} \, .
\label{gen2}
\end{eqnarray}
(The integration over the $\lambda_{\mu}$ around the saddle-point cancels
against the normalization $Z$.) Again, the saddle-point approximation is
used to integrate over $\tau$. The last exponential in Eq.\ (\ref{gen2}) has
a term of order ${\cal O}(1)$ in the exponent and can be omitted. Hence, for
fixed $\sigma$ the equation $i \sigma=  F(E+\tau^{\rm sp})$ determines the
saddle-point  $\tau^{\rm sp} (\sigma ,E)$. By expanding the exponent to
quadratic order in the fluctuations  $\delta \tau$, one can easily verify that
the integral over $\delta \tau$ yields unity. Therefore, we obtain
\begin{eqnarray}
  I\!=\!\! \int\!\! d \sigma\, & & \exp \left\{ {i\over 2} N {\rm trg} (\sigma
  \tau^{\rm sp})-{1\over 2} \sum_{\mu} {\rm trg}  \ln (\lambda_\mu
  ^{\rm sp} -\! E -\! \tau^{\rm sp}) \right\}  \nonumber \\
  & & \times \exp \left\{ -{1\over 2} {\rm tr}\,{\rm trg} \ln
  (1-i\sigma {\bf M}) \right\} ,
\label{gen3}
\end{eqnarray}
where now only the integration over the supermatix $\sigma$ remains
to be done. The saddle-point $\sigma^{\rm sp}$ is found from
the first two terms in the exponent
\begin{equation}
  i \sigma^{\rm sp} {\partial \tau^{\rm sp} \over \partial \sigma} |_{
  \sigma^{\rm sp}}+i \tau^{\rm sp} = F (E+\tau^{\rm sp})
  {\partial \tau^{\rm sp} \over \partial \sigma} |_{\sigma^{\rm sp}} \, .
\label{sigmasp}
\end{equation}
The saddle-point equations for $\tau$ and $\sigma$ together show that
$\tau^{\rm sp} (\sigma^{\rm sp})=0$ and $i \sigma^{\rm sp}=F(E)$.
One observes that the saddle-point equation for $\sigma^{\rm sp}$
is invariant under pseudo-unitary transformations. This implies that
also $\sigma_{\rm G}= T^{-1} \sigma^{\rm sp} T$
is a saddle-point, where $T$ generates pseudo-unitary transformation on
the space of supermatrices. In general $\sigma_{\rm G} \neq \sigma^{\rm
sp}$ and hence the  $\sigma_{\rm G}$ form a manifold of solutions of
the saddle-point equation. We expand the exponential in Eq. (\ref{gen3})
in the vicinity of
the saddle-point solution $\sigma_{\rm G}$, carry out the integral
over the massive modes (which gives unity) and find the result
\begin{equation}
  I=\int \,d\mu (t) \exp \left\{ -{1\over 2} {\rm tr} \, {\rm trg} \,
  \ln [ 1-T^{-1}F(E)T{\bf M}] \right\} \, ,
\label{result1}
\end{equation}
where the integration is now over the manifold of saddle-points.
As usual \cite{blue} one has to give $E$ an imaginary part such that ${\rm
Im} F(E) \sim L$ to guarantee
convergence. Both the structure of the saddle-point manifold
and
the measure $d \mu$ depend only on the symmetry of the ensemble and on
the dimension of the supervectors $\Psi$, $\Psi^\dagger$ in Eq.\
(\ref{genfunc}). In particular, both are independent of the probability density
$P(H)$. On the other hand, the latter specifies the mean
level density $\rho(E)$ and the function $F(E)$ and hence sets the local energy
scale. To see this most clearly one chooses $E$ such that ${\rm Re} F(E)=0$ and
therefore $F(E)=-i \pi \rho(E) L$ \cite{center}. For a symmetric confining
potential $V$ one may take $E$ at the center of the spectrum, $E=0$. Then
\begin{equation}
  I=\int \,d\mu (t) \exp \left\{ -{1\over 2} {\rm tr} \, {\rm trg} \,
  \ln [ 1+i \pi \rho(E) T^{-1} L T {\bf M}] \right\} \, .
\label{result2}
\end{equation}
This proves universality: Our result has the same form as in the Gaussian case
\cite{blue},
the probability density enters only through the local mean level density
$\rho(E)$. On the scale set by $\rho(E)$ all derivatives
of $I$ and hence all correlation functions are independent of the
probability density $P(H)$ and are thus universal.

The proof, presented here for the unitary ensemble, applies equally to the
orthogonal and the symplectic ensemble. In either of the latter cases, the
structure of the supervector $\Psi$ differs from the unitary case. This
structure is reflected in the symmetry properties and dimensions of the
matrices $\sigma$ and $\tau$ and, eventually, of the matrix $T$ which
generates transformations on the saddle-point manifold. However, our
threefold use of the saddle-point approximation is completly independent of
such symmetry properties.

Last we turn to correlation functions which depend on an external
parameter \cite{Altshuler}. One has to distinguish between two cases: The
external perturbation
may either preserve (case (i)) or violate (case (ii)) the symmetry of
the original matrix ensemble. We consider class (i) first and demonstrate
our point for the case of the 2-point level correlation function.
It is defined by $\langle {\rm tr}(E+\epsilon /2 -H -\sqrt{ \alpha /N} H')
{\rm tr}(E-\epsilon /2 -H +\sqrt{ \alpha /N} H') \rangle$  where the energy
difference $\epsilon$ is ${\cal O}(N^{-1})$ and $H'$ is a matrix ensemble
with the same symmetry and the same distribution function as the original
ensemble $H$. The average is over the distributions of both $H$ and $H'$. For
simplicity we restrict ourselves to symmetric distributions. The factor
$\sqrt{N^{-1}}$ appearing
in the definition of the 2-point correlation function
ensures that the corralations decay on the typical scale $\alpha \sim
{\cal O}(1)$. In the generating functional of Eq.\ (\ref{genfunc}) there
now
appears an additional term $\sqrt{\alpha /N} H' L$ which can be included
into the definition of ${\bf M}$ by replacing ${\bf M} \rightarrow
{\bf M}+\sqrt{{\alpha \over N}} H' L$. After averaging over the ensemble $H$
one finds that $I$ is given by
\begin{equation}
  I=\int \!\! d \mu (t) \langle \exp \left\{ -{1 \over 2} {\rm tr}\,{\rm trg}\,
  \ln \left[
  1+i \pi \rho (E) T^{-1}LT {\bf M}\right]
  \right\} \rangle_{H'} .
\end{equation}
Now the logarithm is expanded in powers of $H'$. Due to the factor
$\sqrt{{\alpha / N}}$, taking the ensemble average over $H'$
reduces to calculating the second cumulant, all higher-order cumulants being
small
in comparison by at least a factor $N^{-1/2}$. The final result for $I$ is
\begin{eqnarray}
  \lefteqn{ I=\int d\mu (t) \exp\left\{-{1 \over 2} {\rm tr}\,{\rm trg}\,
  \ln \left[
  1+i\pi \rho (E) T^{-1}LT {\bf M}\right] \right\}} \nonumber \\
  & & \times \exp\left\{ -{\alpha \over N} ({\pi \rho(E)\over 2})^2 \left[
  {\rm trg}
  (T^{-1}LTL)^2  \langle {\rm tr}\, (H')^2 \rangle_{H'} \right] \right\}\! .
\label{Ipar2}
\end{eqnarray}
The distribution over $H'$ enters only trough its second moment. Its value
for a non-Gaussian distribution for $H'$ differs from what one would find for a
Gaussian distribution. However, this difference does not affect the form
of the correlation function and only leads to a rescaling of the parameter
$\alpha$.
Case (ii) is treated along exactly parallel lines and leads to exactly the same
conclusion: Aside from a scaling factor affecting the parameter which governs
symmetry breaking, the form of the correlation function is the same for
Gaussian and non-Gaussian ensembles.

In summary, we have investigated the consequences of non-Gaussian
probability measures within random-matrix theory. We have shown that
in the limit of a large number of levels global and local
properties of the spectrum separate. Global properties like the mean level
density do depend on the form of the measure. Local properties,
in contrast, are independent of the measure. They are determined only by the
symmetry of the ensemble and they are identical to those of the
corresponding Gaussian ensemble. This holds for all three generic
ensembles and for arbitrary form of the measure. Our analytical result
establishes generally and for the first time
that all local random-matrix correlations are independent
of the measure and hence universal.

We thank F.\ von Oppen and J.\ Zuk for
helpful and informative discussions.


\begin{references}
\bibitem{Kravtsov} In particular, $V(\lambda )$ is independent of $N$ and
increases at least like $|\lambda |$ for $\lambda \rightarrow \infty$.
\bibitem{center} This choice is always possible. By assumption, the
potential $V(H)$ confines the spectrum
to some finite interval, hence for some $E$ within the spectrum we must
have $V'(E)=0$. On the other hand, the resolvent for $\epsilon \rightarrow 0$
has the property \cite{Brezin1} that $F(E+i \epsilon L)= V'(E)/2-i\pi \rho(E)
L$.
\end{references}
\end{document}